\documentclass[journal]{IEEEtran}
\usepackage[T1]{fontenc}
\usepackage{amsmath,graphicx}

\usepackage[affil-it]{authblk}

\usepackage{url}
\usepackage{amsmath,amssymb,amsfonts}
\usepackage{algorithmic}
\usepackage{graphicx}
\usepackage{textcomp}
\usepackage{xcolor}
\usepackage[ruled,vlined]{algorithm2e} 
\newcommand{\pd}{\textcolor{black}}

\usepackage{enumitem}
\usepackage{siunitx}

\begin{document}
\title{A Novel Fast 3D Single Image Super-Resolution Algorithm}

\author{	
	Nwigbo Kenule Tuador, 
	Duong Hung Pham,
	Jérôme Michetti,
	Adrian Basarab,
	Denis Kouamé
}

	\affil{{\small University of Toulouse, IRIT, CNRS UMR 5505, Universit\'e Paul Sabatier, Toulouse, France}
 }
\maketitle
\begin{abstract}
This paper introduces a novel computationally efficient method of solving the 3D single image super-resolution (SR) problem, i.e., reconstruction of a high-resolution volume from its low-resolution counterpart. The main contribution lies in the original way of  handling simultaneously the associated decimation and blurring operators, based on their underlying properties in the frequency domain. In particular, the proposed decomposition technique of the 3D decimation operator allows a straightforward implementation for Tikhonov regularization, and can be further used to take into consideration other regularization functions such as the total variation, enabling the computational cost of state-of-the-art algorithms to be considerably decreased. Numerical experiments carried out showed that the proposed approach outperforms existing 3D SR methods.

\end{abstract}

\section{Introduction}
3D \pd{single image super-resolution (SR) is an active subject of study in medical imaging field with various practical applications such as anatomy structure exploration \cite{Greenspan2008}, brain magnetic resonance imaging analysis \cite{Pham2019}, cardiac applications \cite{Oktay2016} or dental cone beam computed tomography (CBCT) \cite{Hatvani2019}. The objective of 3D-SR algorithms is to} reconstruct a 3D high-resolution (HR) image from its observed low-resolution (LR) counterpart, thus allowing for better visualization and interpretation of the image content. \pd{A large number of SR approaches have been proposed in the literature, among which the most popular fall in two categories: model- and learning-based techniques. The latter are usually addressed by convolutional neural networks. However, despite their accuracy, training plays a key role for such approaches, by connecting LR images to their corresponding HR images (see, e.g., \cite{Pham2019,Oktay2016,Hatvani2019}). Unfortunately, learning-based SR methods require large training datasets which, depending on the application, may not always be available. Alternatively, model-based methods express the SR problem as an image reconstruction process, consisting in estimating the image that minimizes a cost function based on a given image degradation model and on prior knowledge about the volume to estimate. Total variation (TV) \cite{Toma2014}, low-rank \cite{Shi2015} or sparsity in a given domain, such as for instance wavelet domain \cite{Zhang2018} are among the most used regularization functions. The degradation model usually combines blurring and decimation operators in order to account for loss of spatial resolution. However, a major drawback of these methods when applied to 3D volumes is their computational complexity and the fact that they rely upon volume unfolding to convert the 3D volumes into 2D matrices or 1D vectors, thus losing the information of voxel neighborhood. To avoid unfolding the volumes, a specific class of 3D SR methods have been recently introduced, based on tensor theory \cite{Kanatsoulis2018, Hatvani2019a,Prevost2020}. The main principle of tensor-based SR approaches is to decompose the volume of interest considered as a 3D-tensor into the sum of elementary tensors, using, for instance, the canonical polyadic decomposition, followed by truncation of the decomposition. Despite their interest and significant computational efficiency, the flexibility of the tensor-based SR methods is mitigated by the fact that no specific regularization is taken into consideration.}

To overcome the limitations of existing approaches, i.e., the need of training databases for deep learning, the computational load of reconstruction-based methods and the lack of regularization functions for tensor-based techniques, this paper introduces an extension of the 2D SR method in \cite{Zhao2016} to 3D. The main contribution of this work lies in exploiting an interesting property of the 3D decimation operator in the Fourier domain, that allows SR algorithms to be implemented analytically for Tikhonov regularization and efficiently for other regularizations such as total variation through an algorithm of alternating direction method of multipliers.

\vspace{-0.25cm}

\section{Background}
\label{sec:back}
\subsection{Image degradation model}
Although the developed technique is general and could be applied to different medical imaging modalities, its interest is illustrated herein in 3D dental CBCT. CBCT induces unavoidable degrading factors such as blurring, down-sampling and noise resulting in the following model: 
\begin{equation}
\textbf{y}=\textbf{D}\textbf{H}\textbf{x}+\textbf{n},
\label{equ1}
\end{equation}
\noindent where \textbf{y} $\in\mathbb{R}^{N_{l} \times 1}$ $(N_{l}=m_{l}\times n_{l}\times s_{l})$ and $\textbf{x}\in\mathbb{R}^{N_{h} \times 1}$ $(N_{h}=m_{h}\times n_{h}\times s_{h})$  are respectively \pd{vectorized versions of  LR image of size $m_{l} \times n_{l} \times s_{l}$ and 
HR image of size $m_{h} \times n_{h} \times s_{h}$, obtained by ordering their voxels lexicographically}. The 3D HR image is altered by a decimation operator $\textbf{D}\in\mathbb{R}^{N_{l} \times N_{h}}$ with an integer rate $d = dr \times dc \times ds$, i.e., $N_{h} = N_{l} \times d$. $\textbf{H}\in \mathbb{R}^{N_{h}\times N_{h}}$ is the point spread function (PSF), and  $\textbf{n}\in\mathbb{R}^{N_{l} \times 1}$ is an additive white Gaussian noise. The decimation rates $d_{r}, d_{c}$ and $d_{s}$ correspond to the pixel resolution loss in each spatial direction, satisfying  
$m_{h} = m_{l} \times d_r $, $ n_{h} = n_{l} \times d_c$ and $s_{h} = s_{l} \times d_s.$ Based on the common assumption of cyclic convolution, the blurring operator $\textbf{H}$ and its conjugate transpose $\textbf{H}^{H}$ can be decomposed as \cite{Hansen2006}:
\begin{equation}\label{equ2}
\textbf{H} = \textbf{F}^H \mathbf{\Lambda} \textbf{F} \quad \mathrm{and}\quad \textbf{H}^{H} = \textbf{F}^H \mathbf{\Lambda}^{H} \textbf{F},
\end{equation}
\noindent where $\mathbf{\Lambda}\in\mathbb{C}^{N_{h} \times N_{h}} $ is a diagonal matrix \pd{whose elements are the 3D Fourier transform of the zero-padded PSF, i.e., of the first column of \textbf{H}}, and $\textbf{F} \in\mathbb{C}^{N_{h} \times N_{h}}$ is the unitary 3D Fourier transform matrix having the property $\textbf{F}^H=\textbf{F}^{-1}$. While the decimation operator  $\textbf{D}\in\mathbb{R}^{N_{l} \times N_{h}}$ downsamples a 3D image, its conjugate transpose $\textbf{D}^H \in \mathbb{R}^{N_{h} \times N_{l}}$ is assumed to interpolate the image with zeros \cite{Zhao2016}. Thus, $\textbf{DD}^H=\textbf{I}_{N_{l}}$ and, by notation, $\underline{\textbf{D}}\triangleq \textbf{D}^H\textbf{D}$. 

\vspace{-0.25cm}

\subsection{Super-resolution problem formulation}
The estimation of the HR image from the observed LR image is usually expressed as a maximum \textit{a posteriori} (MAP) estimator that allows including \textit{a priori} information about the image to estimate. Based on the image formation model in \eqref{equ1} the MAP estimator can be expressed as:
\begin{equation} 
\hat x= {\underset{\textbf{x}}{\mathrm{argmin}}}\frac{1}{2}\underbrace{||\textbf{y}-\textbf{DHx}||_{2}^{2}}\ + \lambda \underbrace{\phi (\textbf{Lx}) },
\label{equ3}
\end{equation}
\noindent where the first term is the data fidelity term, and the second term is a regularization function related to the \textit{a priori} statistical assumption on $\textbf{x}$. $\textbf{L}$ may be the identity matrix if the prior distribution is considered in the spatial domain, or any basis or dictionary used to transform the image into a different domain. $\mathbf{\lambda} > 0$ is a hyperparameter that dictates the compromise between data fidelity and regularization terms.

\vspace{-0.2cm}

\section{\pd{Proposed fast 3D super-resolution algorithm}}

The high dimension of the images represents an important challenge in minimizing the function in \eqref{equ3}. This is partially mitigated, for the blurring operator, by the assumption of cyclic convolution that, following \eqref{equ2}, allows its implementation by point-wise multiplications in the 3D Fourier domain. However, the decimation operator $\textbf{D}$ does not have the same property and can thus not be handled in the same way as $\textbf{H}$. The main aim of this work is to exploit a different property of $\textbf{D}$ to enable an efficient implementation of numerical optimization algorithms able to solve \eqref{equ3}. In the following subsections, the proposed method is shown in the case of two different regularization functions. 


\vspace{-0.3cm}

\label{sec:3D_SISR}
\subsection{Tikhonov regularization}

\pd{We first consider the simple case of the Tikhonov regularization, resulting into the following function to minimize:}
\begin{equation} 
\hat x= {\underset{\textbf{x}}{\mathrm{argmin}}}\frac{1}{2}||\textbf{y}-\textbf{DHx}||_{2}^{2} + \lambda||\textbf{x}-\overline{\textbf{x}}||_{2}^{2},
\label{equ5}
\end{equation} 
\noindent where $\overline{\textbf{x}}\in\mathbb{R}^{N_{h} \times 1}$ is an approximation of the SR image to estimate.
Equation \eqref{equ5} has a closed form solution:

\begin{equation}
\hat{\textbf{x}}=(\textbf{H}^H\underline{\textbf{D}}\textbf{H}+2\lambda \textbf{I}_{N_{h}})^{-1}(\textbf{H}^H\textbf{D}^H\textbf{y}+2\lambda \overline{\textbf{x}}),
\label{equ6}
\end{equation}

\noindent or, in the Fourier domain, using \eqref{equ2}:
\begin{eqnarray}
\hat{\textbf{x}}=\textbf{F}^H(\mathbf{\Lambda}^H\textbf{F}\underline{\textbf{D}}\textbf{F}^H\mathbf{\Lambda}+2\lambda \textbf{I}_{N_{h}})^{-1}\textbf{F}\textbf{k},
\label{equ7}
\end{eqnarray}
\noindent where $\textbf{k} = (\textbf{F}^H\mathbf{\Lambda}^H\textbf{F}\textbf{D}^H\textbf{y}+2\lambda \overline{\textbf{x}})$. None of the analytical solutions in \eqref{equ6} and \eqref{equ7} can be used in practical 3D applications. In \eqref{equ6}, the matrices to manipulate and to invert are of size $N_{h} \times N_{h}$. In \eqref{equ7}, while the matrix multiplications with $\mathbf{\Lambda}$ is straightforward (element-wise multiplications), it is not with $\textbf{D}$ that is not diagonal. However, the generalization to 3D, proposed herein, of the theoretical result presented in \cite{Zhao2016} for 2D images, allows the following decomposition:
\begin{equation}
\textbf{F}\underline{\textbf{D}}\textbf{F}^H = \frac{1}{ds}(\textbf{J}_{ds}\otimes \textbf{I}_{s_{l}})\otimes \frac{1}{dc}(\textbf{J}_{dc}\otimes \textbf{I}_{n_{l}})\otimes \frac{1}{dr}(\textbf{J}_{dr}\otimes \textbf{I}_{m_{l}}),
\label{equ8}
\end{equation}
\noindent where \pd{$\textbf{J}_{u}\in \mathbb{R}^{u \times u}$ is the $u\times u$ matrix of ones, $\textbf{I}_{v}\in \mathbb{R}^{v \times v}$ is the $v\times v$ identity matrix and $\otimes$ is the Kronecker product. Using this result, \eqref{equ7} may be rewritten as:}
\begin{equation}
\hat{\textbf{x}}=\textbf{F}^H(\frac{1}{r}\underline{\mathbf{\Lambda}}^H\underline{\mathbf{\Lambda}}+2\lambda \textbf{I}_{N_{h}})^{-1}\textbf{F}\textbf{k},
\label{equ9}
\end{equation}
 where $\underline{\mathbf{\Lambda}}\in\mathbb{C}^{N_{l} \times N_{h}}$ is defined as:
\begin{equation*}
\underline{\mathbf{\Lambda}}=\underbrace{\left[(\textbf{1}_{ds}^T\otimes \textbf{I}_{s_{l}})\otimes(\textbf{1}_{dc}^T\otimes \textbf{I}_{n_{l}})\otimes (\textbf{1}_{dr}^T\otimes \textbf{I}_{m_{l}})\right]}_{\in~\mathbb{C}^{N_{l} \times N_{h}} }\mathbf{\Lambda},
\end{equation*}

\noindent where $\textbf{1}_{w}^T\in\mathbb{R}^{1 \times w}$ is a vector of ones.
\eqref{equ9} can be further simplified using the matrix inversion lemma \cite{Hager1989}: 
 

\vspace{-0.4cm}
\begin{equation}
    \hat{\textbf{x}} = \frac{1}{2\lambda}\textbf{k}- \frac{1}{2\lambda}\textbf{F}^H\underline{\mathbf{\Lambda}}^H(2\lambda r\textbf{I}_{N_{l}}+\underline{\mathbf{\Lambda}}\underline{\mathbf{\Lambda}}^H)^{-1}\underline{\mathbf{\Lambda}}\textbf{Fk}.
    \label{equ12}
\end{equation}

It is worth mentioning that the numerical implementation of \eqref{equ12} requires only one 3D fast Fourier transform and one 3D fast inverse Fourier transform, in addition to element-wise multiplications, reducing its computational load drastically. Compared with the original solution \eqref{equ6}, the computational complexity order of this fast solution plummets from $\mathcal{O}(N_{h}^3)$ to $\mathcal{O}(N_{h}\log N_{h})$.
\eqref{equ12} is referred to as a 3D-fast super resolution  (3D-FSR) and denoted by 3D-FSR-$l_2-l_2$ in the sequel.

\vspace{-0.45cm}
\subsection{Total variation regularization}

In this subsection, we show how \eqref{equ12} can be used to handle other regularization terms. For illustration purpose, we consider hereafter the total variation, expressed as:
\begin{equation}
    \phi(\textbf{Lx}) = \|x\|_{\textbf{TV}} = \sqrt{\|\textbf{D}_h\textbf{x}\|^2 + \|\textbf{D}_v\textbf{x}\|^2 + \|\textbf{D}_s\textbf{x}\|^2},
\end{equation}
\noindent where $\textbf{D}_h$, $\textbf{D}_v$ and $\textbf{D}_s$ are standard finite difference operators in each spatial direction. Note that the problem to solve in this case is the sum between the data fidelity term in \eqref{equ3} and the total variation term above. One common way to solve such problems is to use the alternate direction method of multipliers (ADMM) \cite{Boyd2010}. To this end, the function minimization is first turned into the following constrained optimization problem: 
\vspace{-0.2cm}
\begin{eqnarray}
(\hat{\textbf{x}},\hat{\textbf{u}}) = {\underset{\textbf{x},\textbf{u}}{\mathrm{argmin}}}  & & \frac{1}{2}||\textbf{y}-\textbf{DHx}||_{2}^{2}+\lambda \phi (\textbf{u})  \nonumber \\
  \textrm{subject to} & & \textbf{Lx} =\textbf{u},
  \label{eq_ADMM}
\end{eqnarray}
\vspace{-0.1cm}
 \noindent whose associated augmented Lagrangian function is: 
 \vspace{-0.2cm}
 \begin{equation}
\mathcal{L}(x,u,d)= \frac{1}{2}||\textbf{y} - \textbf{DHx}||_{2}^{2} + \lambda \phi(\textbf{u}) + \frac{\mu}{2}||\textbf{Lx} - ( \textbf{u} - \textbf{d})||_{2}^{2},
 \end{equation}

\noindent where $\textbf{L} = \left[\textbf{D}_h, \textbf{D}_v, \textbf{D}_s\right]^T \in \mathbb{R}^{3N_h \times N_h}$ and  $\textbf{u} = \left[\textbf{u}_h, \textbf{u}_v, \textbf{u}_s\right]^T \in \mathbb{R}^{3N_h \times 1}$. ADMM minimizes $\mathcal{L}(\textbf{x},\textbf{u},\textbf{d})$ by solving iteratively subproblems over each variable $\textbf{x}$, $\textbf{u}$ and scaled dual variable $\textbf{d}$:
\vspace{-0.6cm}

\begin{equation}
\begin{array}{l}
\text{For}\; i = 0,\ldots\\
\left\lfloor
\begin{array}{l}
\textbf{x}^{i+1} =  {\underset{x}{\mathrm{argmin}}}\ \frac{1}{2}||\textbf{y}-\textbf{DHx}||_{2}^{2}+ \frac{\mu}{2}||\textbf{Lx}-( \textbf{u}^i-\textbf{d}^i)||_{2}^{2} \\
\textbf{u}^{i+1} = {\underset{u}{\mathrm{argmin}}} \ \lambda \phi(\textbf{u})+ \frac{\mu}{2}||\textbf{Lx}^{i+1}-( \textbf{u}-\textbf{d}^i)||_{2}^{2} \\
\textbf{d}^{i+1} = \textbf{d}^i+ (\textbf{Lx}^{i+1}- \textbf{u}^{i+1})
\end{array}
\right.
\end{array} 
\label{eq_framework}
\end{equation}

\vspace{-0.1cm}
One may note that the first step, i.e., the minimization over $\textbf{x}$, can be solved using the method proposed in \eqref{equ12}. The resulting algorithm able to solve efficiently the SR problem with TV regularization is given in Algorithm \ref{algo:l2TV}. Note that the derived solution in \eqref{equ12} plays a central role in this algorithm, which considerably decreases the computational burden of existing SR methods, as shown in the results section \ref{sec:numerical_results}.

\begin{algorithm}
\caption{3D-FSR-$l_2-TV$}
\KwIn{\textbf{y}, \textbf{H}, $\lambda$, r} 
\textbf{Initialize:} Set i = 0, choose $\mu > 0, \textbf{d}^0, \textbf{u}^0$
\begin{enumerate}[label={\arabic*.}]
    \item Factorize: $\textbf{H} = \textbf{F}^H \mathbf{\Lambda} \textbf{F}$
\item Compute $\underline{\mathbf{\Lambda}}$: $\underline{\mathbf{\Lambda}}= \left[(1_{ds}^T\otimes \textbf{I}_{s_{l}})\otimes(1_{dc}^T\otimes \textbf{I}_{n_{l}})\otimes (1_{dr}^T\otimes \textbf{I}_{m_{l}})\right]\mathbf{\Lambda}$\\
\item Factorize:
$\textbf{D}_h = \textbf{F}^H \mathbf{\Sigma}_h \textbf{F}$, $\textbf{D}_v = \textbf{F}^H \mathbf{\Sigma}_v \textbf{F}$, $\textbf{D}_s = \textbf{F}^H \mathbf{\Sigma}_s \textbf{F}$

\item Compute: $\mathbf{\Gamma} = (\mathbf{\Sigma}_h^H\mathbf{\Sigma}_h+\mathbf{\Sigma}_v^H\mathbf{\Sigma}_v+\mathbf{\Sigma}_s^H\mathbf{\Sigma}_s)^{-1}$\\
\item 
 \While{stopping criterion is not satisfied}{   
   \begin{enumerate}[label={\alph*.},leftmargin=0pt,align=left]
    \item Update \textbf{x} using \eqref{equ12}:\\
$\rho_h = \textbf{u}_h^i - \textbf{d}_h^i$, $\rho_v = \textbf{u}_v^i - \textbf{d}_v^i$, $\rho_s = \textbf{u}_s^i - \textbf{d}_s^i;$\\
$\mathbf{\Theta} = \textbf{D}_h^H\rho_h+\textbf{D}_v^H\rho_v+\textbf{D}_s^H\rho_s$\\
 Determine \textbf{k}: $\textbf{k} = \textbf{F}^H\mathbf{\Lambda}^H\textbf{FD}^H\textbf{y}+\mu \Theta$ \\
 Determine 3D FFT of \textbf{k}: 
$\textbf{Fk} = \mathbf{\Lambda}^H\textbf{FD}^H\textbf{y}+\mu \textbf{F}\Theta$ \\
 Entrywise product: $\textbf{x}_f=(\mathbf{\Gamma}\underline{\mathbf{\Lambda}} ^H(\mu  r\textbf{I}_{N_{l}}+\underline{\mathbf{\Lambda}}\mathbf{\Gamma} \underline{\mathbf{\Lambda}}^H)^{-1}\underline{\mathbf{\Lambda}}\mathbf{\Gamma})\textbf{Fk}$

$\textbf{x}^{i+1} =  \frac{1}{\mu}(\textbf{F}^H\mathbf{\Gamma} \textbf{Fk}-\textbf{F}^H\textbf{x}_f)$
\item Update \textbf{u} using the soft-thresholding operator:\\
$\nu = \left[\textbf{D}_h\textbf{x}^{i+1}+\textbf{d}_h^i,\textbf{D}_v\textbf{x}^{i+1}+\textbf{d}_v^i,\textbf{D}_s\textbf{x}^{i+1}+\textbf{d}_s^i\right]$\\
$\textbf{u}^{i+1}\left[j\right] = \max \left\{0,||\nu\left[j\right]||_2 - \frac{\lambda}{\mu} \right\}\frac{\nu \left[j\right]}{||\nu \left[j\right]||_2}$
\item Update: $\textbf{d}^{i+1} = \textbf{d}^i + (\textbf{L}\textbf{x}^{i+1} - \textbf{u}^{i+1});$\\
     \end{enumerate}           
 } 
\end{enumerate}
\KwOut{$\hat{\textbf{x}} = \textbf{x}^i.$}
\label{algo:l2TV}
\end{algorithm}

\vspace{-0.5cm}

\begin{table}[!ht]
\caption{Quantitative results obtained with ADMM and 3D-FSR-$l_2-l_2$.}
\begin{center}
\begin{tabular}{ |p{2.2cm}|p{2cm}|p{2cm}|}
\hline

 Method & PSNR(dB) & Time (sec.)   \\ 
 \hline\hline
 ADMM & \textbf{39.25} & 397.93 \\
 \hline
 3D-FSR-$l_2-l_2$& \textbf{39.25}&  \textbf{2.48} \\
 
\hline
\end{tabular}
\label{tab1}
\end{center}
\end{table}


\begin{table}[!htb]
\caption{Parameters associated with each method in $l_2-TV$ case}
\begin{center}
\begin{tabular}{ |p{3.3cm}|p{1.5cm}| p{2.1cm}|  }
\hline
 \hspace{1.3cm}LRTV & \hspace{0.2cm}TF-SISR & 3D-FSR-$l_2-TV$  \\ 
 \hline\hline
 $\lambda_{TV} = 0.02$  &  $ n_{TF} = 20$ & $\lambda = 0.06$\\
 
 $\lambda_{R} = 0, \rho = 0.05$ & $F = 900 $ & $ \mu = 0.1 $ \\
 
 $ n_{grad} = 100, dt = 0.05$ & $ \epsilon = 1 $ & $ n_{3D-FSR} = 30 $ \\ 
\hline
\end{tabular}
\label{tab2}
\end{center}
\end{table}

\vspace{0.2cm}

\begin{table}[!h]
\caption{Quantitative results for $TV$ regularization.}
\begin{center}
\begin{tabular}{ |p{2.3cm}|p{2.3cm}|p{2.3cm}|  }
\hline
 Output & PSNR (dB) & Time (sec.)  \\ 
 \hline\hline
 CBCT & 23.21 & - \\ 
 \hline
 LRTV & 24.23 & 1104.12 \\
 \hline
 TF-SISR & 24.52 & 38.51 \\
 \hline
 3D-FSR-$l_2-TV$ &  \textbf{25.37} & \textbf{223.61} \\ 
\hline
\end{tabular}
\label{tab3}
\end{center}

\end{table}

\vspace{-0.1cm}
\section{Numerical Results}
\label{sec:numerical_results}

This section regroups numerical experiments conducted on dental datasets to illustrate the benefits of using the proposed method with Tikhonov and TV regularizations. Two volumes have been acquired on an extracted tooth, using a $\mu$CT Perkin Elmer Quantum FX and a CBCT Carestream 81003D scanner. $\mu$CT provides high-resolution images (voxel scale of $40 \times 40 \times 40$ \SI{}{\micro\metre}$^3$ and a linewidth resolution of \SI{40}{\micro\metre}) at the cost of high irradiation doses that prevent it from clinical applications. In contrast, CBCT is well-accepted in dental applications, but its spatial resolution (linewidth resolution \SI{500}{\micro\metre} and a voxel size of $80 \times 80 \times 80$ \SI{}{\micro\metre}$^3$) is not enough for, e.g., endodontics, the dental specialty concerned with the study and treatment of the dental pulp. The functionality of the SR algorithms evaluated is measured both visually and quantitatively in terms peak signal-to-noise ratio (PSNR) and computational time (in seconds). All numerical experiments were carried out on a computer with Windows 10 Pro, Intel(R) Core (TM) i5-8500 CPU @3.00GHz and 16 GB RAM with Matlab R2019a
\footnote{The Matlab code of the proposed algorithms is available at https://www.irit.fr/~Adrian.Basarab/img/Demo3DFSR.zip}.



\vspace{-0.3cm}

\subsection{Results with Tikhonov regularization}
A synthetic low resolution image was created from the $\mu$CT volume, by blurring it with a 3D Gaussian smoothing kernel of size $9 \times 9 \times 9$ with standard deviation $\sigma=3$ in each spatial dimension, down-sampling it with decimation factors $d_{r}=d_{c}=d_{s}=2$ and contaminating it by an additive white Gaussian noise characterized by a blurred signal-to-noise ratio (BSNR) of 30 dB. The high resolution resolution volume was estimated using \eqref{equ12}, and by minimizing \eqref{equ5} iteratively using a straightforward ADMM algorithm. From Table \ref{tab1}, one may notice that the two techniques produce similar reconstruction qualities in terms of PSNR, but the proposed technique performs significantly faster than ADMM.

\begin{figure*}[!h]
 \centering
  \includegraphics[width=\linewidth,height=5cm]{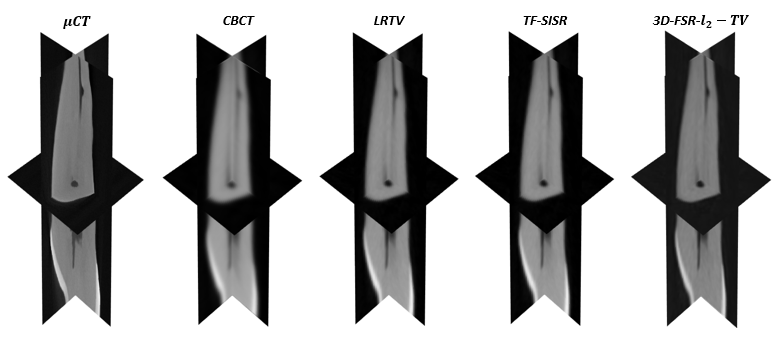}
  \caption{Slice planes extracted from the resulting volumes obtained by $\mu$CT, CBCT, LRTV, TF-SISR and 3D-FSR-$l_2-TV$, respectively.}
  \label{fig:boat2}
\end{figure*}

\begin{figure}[!h]
    \centering
  \includegraphics[width=0.75\linewidth]{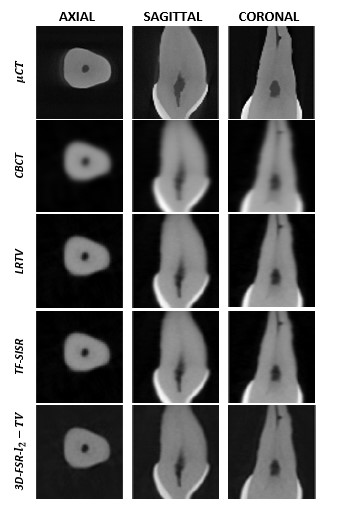}
  \caption{Axial, saggital and coronal slices extracted from the resultant volumes given by $\mu$CT, CBCT, LRTV, TF-SISR and 3D-FSR-$l_2-TV$. The root canal corresponds to the dark region inside the tooth.}
  \label{fig:boat1}
\end{figure}

\vspace{-0.3cm}

\subsection{Results with total variation regularization}
The second series of results aims at validating Algorithm \ref{algo:l2TV}. In this case, the CBCT volume was considered as the LR images to be enhanced and the $\mu$CT volume as the HR ground truth. The PSF was unknown in this case and estimated from the CBCT volume by approximating it with a 3D Gaussian function, as suggested in \cite{Hatvani2019a}. The estimated standard deviations in the three spatial dimensions were $\sigma_1 = 5.8$, $\sigma_2 = 5.3$ and $\sigma_3 = 0.9$. Algorithm \ref{algo:l2TV} was compared to two different 3D SR methods from the literature: (i) the method in \cite{Shi2015}, denoted by LRTV, a reconstruction approach that unfolds the volumes to 2D matrices, and (ii) the tensor-based (TF-SISR) method in \cite{Hatvani2019a}. Note that LRTV originally uses both low rank and TV regularizations, but was used herein only with TV to ensure fair comparison. The hyperparameters of the three methods, regrouped in Table \ref{tab2}, have been tuned by cross-validation to maximize the PSNR computed between the restored and $\mu$CT volumes. The quantitative results are given in Table \ref{tab3}. They show a slight PSNR improvement enabled by the proposed method and a significant gain in computational time compared to LRTV. Note that TF-SISR is also computational effective but, in contrast to the proposed approach, offers no flexibility in choosing the desired regularization function. The $\mu$CT, CBCT and restored volumes are shown in Figs. \ref{fig:boat2} and \ref{fig:boat1}. One may notice the ability of the SR algorithms to enhance the CBCT images, and in particular to enhance the canal root.


\vspace{-0.25cm}

\section{Conclusions}
\vspace{-0.2cm}
In this paper, a new model-based technique for 3D SR problem was introduced, effectively computing the 3D super resolved image by utilizing specific attributes of the 3D decimation and blurring operators in the Fourier domain. Indeed, these attributes enabled the 3D SR proposition to cope with standard priors like Tikhonov and TV regularizations, leading to fast implementation schemes. This novel method offered more computational efficiency and flexibility compared to existing reconstruction- and tensor-based techniques respectively. Further work can extend the proposed method to 3D multiframe super-resolution or to non-Gaussian noise.

\vspace{-0.25cm}
\section{Compliance with Ethical Standards}
\label{sec:ethics}
\vspace{-0.3cm}
This is a numerical simulation study for which no ethical approval was required.
 
\vspace{-0.35cm} 
 \section{Acknowledgments}
\label{sec:acknowledgments}
\vspace{-0.3cm}
This work has been supported by the PTDF overseas scholarship under the government of Nigeria.


\vspace{-0.3cm}

\bibliographystyle{IEEEtran}

\footnotesize
\bibliography{ISBI2020.bib}
\end{document}